\documentclass{kluwer}    

\newdisplay{guess}{Conjecture}
\usepackage{epsfig}

\begin{document}
\begin{article}

\begin{opening}
\title{Flexible Automatic Scheduling For Autonomous Telescopes: The MAJORDOME}
\author{Mathieu \surname{ Bringer}, Michel \surname{ Bo\"er}\email{boer@cesr.fr}, Cedric \surname{Peignot}}
\institute{$^1$Centre d'Etude Spatiale des Rayonnements
(CESR - CNRS)
\\9, ave. du Colonel Roche
\\BP 4346
\\F 31028 Toulouse Cedex 4
\\France}
\author{G\'erard \surname{Fontan}, Colette \surname{Merc\'e}}
\institute{Laboratoire d'Automatique et d'Analyse des
Syst\`emes (LAAS - CNRS)
\\7, ave. du Colonel Roche
\\F 31077 Toulouse Cedex 4
}

\runningauthor{M. Bringer, et al.} \runningtitle{The
MAJORDOME}
\date{November 22, 2000}

\begin{abstract}
We have developped a new method for the scheduling of
astronomical automatic telescopes, in the framework of the
autonomous TAROT instrument. The MAJORDOME software can
handle a variety of observations, constrained, periodic,
etc., and produces a timeline for the night, which may be
modified at any time to take into account the specific
conditions of the night. The MAJORDOME can also handle
target of opportunity observations without delay.
\end{abstract}

\end{opening}

\section{Introduction}

With the advent of automatic telescopes and the use of large
facilities in service mode, efficient automatic scheduling of the
night observations and operations is becoming more important. The
problem may be summarized as how to observe as much as
possible sources, minimizing the loss of observing time, ensuring
optimal observations as far as possible, including several levels
of priority and constraints, calibrating observations, and taking
into account that night operations may be interrupted
 at any time by bad weather, failure of
any device, or by an unexpected event. In the usual configuration
an observer is allocated a given amount of time (hours or nights)
which are usually pre-allocated at a given (block of) nights.

A mixed approach has been introduced for the ESO VLT and NTT
\cite{cha98} with the use of the concept of "observation blocks",
defined as programmable units. Each accepted proposal is
described in terms of "observation blocks" which are then handled
by 3 schedulers, in charge respectively of the long term, the
medium term and the short term timeline, with a degree of
reactivity growing from the long to the short term. The
complexity of the system comes primarily from the need to cope
with observations performed by guest as well as service
observers, and in addition from the multi-telescope nature of the
VLT. In a somewhat different approach, Brezina (1998) defines
individual groups of observations sent to the telescope via a
centralized system, the "Principal Astronomer", with pre-allocated
priorities and time constraints. The system chooses the best
sequence which optimizes an objective function. However, TOO
observations as well as changes in schedule are hardly taken into
account by this system.

The primary goal of the {\it T\'elescope \`a Action Rapide pour
les Objets Transitoires} (Rapid Action Telescope for Transient
Objects, hereafter TAROT, \cite{boer99}\cite{boer01}) is the
simultaneous observation of cosmic Gamma-Ray Bursts (hereafter
GRB) at gamma-ray and visible wavelengths. For that purpose,
TAROT is able to move to any location on the sky upon the receipt
of a position transmitted via the INTERNET by the GRB Coordinate
Network \cite{bar97}.
GRB counterpart
observations use less than 10\% of the telescope time, including
the time needed to follow the source after the immediate
observation, both the same, and the following nights (typically 8
- 10h every 10 nights) . During the 90\% remaining useful time
TAROT observes various categories of sources, e.g. AGN, X-ray
binaries, detection of exo-planets, etc., usually connected with
celestial variability.

One of the peculiarities of TAROT of relevance for the topic of
telescope scheduling is its reactivity to any unprogrammed event,
and the absence of people in charge of the telescope service, at
night, as well as during the day.

For that reason we developed a scheduling system, the MAJORDOME,
in charge of the telescope schedule and which is the interface
between the users and the satellites which give sources to point
and the Telescope Control System (TCS). It has for mission to
handle the user requests (routine program), as well as the
un-programmed events, usually the bad weather, and sometime GRB
alerts.

 In this paper we present our framework for telescope scheduling
activities and its first implementation called the MAJORDOME. In
the next section we describe the TAROT telescope since it was the
instrument used to test the method presented here. In section 3
we present the requirements we imposed on the scheduling task and
the measures we used to control the quality of the solution. In
section 4 we develop the present implementation and the relations
between the MAJORDOME and the other TAROT modules. In section 5
the results obtained both on simulated and real data are
presented. Section 6 is devoted to the discussion, the conclusion,
and the perspectives for the second version of the MAJORDOME
which is currently under development.

\section{The TAROT autonomous telescope}

TAROT-1 is a fully autonomous 25 cm aperture telescope. Its
$2\deg$ field of view matches well the HETE uncertainty in
localization of the sources (particularly for the HETE-1 WXM,
since TAROT-1 was designed before its unsuccessful launch). Table
1 summarizes the present main technical characteristics of
TAROT-1. Note that TAROT performs only imaging.

\begin{table}
\caption[]{Main technical characteristics of TAROT-1}
\label{TAROT1}
\[
         \begin{array}{ll}
            \hline
            \noalign{\smallskip}
\mathrm{Aperture} & 25\,\mathrm{cm} \\
\mathrm{Field\:of\:view} & 2 \deg \times 2 \deg \\
\mathrm{Optical\:resolution} & 20\,\mu\mathrm{m} \\
\mathrm{Mount\:type} & \mathrm{equatorial} \\
\mathrm{Axis\:speed\:} (\alpha\: and\: \delta ) &
                    \mathrm{adjustable,\:up\:to\:}80\deg / \mathrm{s} \\
\mathrm{CCD\:type} & Thomson THX\:7899 \\
\mathrm{CCD\:size} & 2082 \times 2072\,\mathrm{pixels} \\
\mathrm{Pixel\:size} & 15\,\mu\mathrm{m} \\
\mathrm{CCD\:readout\:noise} & \approx 14\,\mathrm{e}^- \\
\mathrm{Readout\:time} & 2\,\mathrm{s} \\
\mathrm{Filter\:wheel} & 6\:\mathrm{pos.:\,Clear,\,
V,\,R,\,I,\,B}+\mathrm{V,\,R}+\mathrm{I}^\mathrm{a} \\
 \noalign{\smallskip}
            \hline
         \end{array}
      \]
\begin{list}{}{}
\item[$^{\mathrm{a}}$] Filters B+V and R+I are broad band filters
covering the spectral range of respectively the Cousin B and V,
and R and I filters.
\end{list}
\end{table}

TAROT-1 is fully autonomous, i.e. there is no human intervention
either night and day, excepted for the weekly replacement of the
DAT archive medium, and low level maintenance, e.g. to verify the
camera vacuum (once per month), cleaning of the optics (6 months)
etc., or low level diagnostic or emergency intervention in case
of a failure (as an illustration this happened when one of the
drives of the sliding roof failed in winter).

 \begin{figure}[h!]
  \caption{The software modules of TAROT and their interactions. In the
  case of GRBs, requests are sent to the MAJORDOME by the GCN in immediately scheduled.
  Routine observations
  are requested through the "Observation Requestor" (with a web interface),
  and scheduled by the MAJORDOME according to the rules described in this paper.
  As soon as the image has been acquired, it is processed by the TAITAR software, and
  accessible to the users via a web interface.}
  \end{figure}

Figure 1 summarizes the different software modules and their
connections. All are functionally independent, and the the
communications are made through TCP/IP socket processes.
Observation requests are sent to the MAJORDOME via a web
interface, and handled as described in the following sections.
Should a GRB occurs, its coordinates are received from the GCN,
and the MAJORDOME immediately schedules the observation: in that
case, the typical slewing times range from 1 to 2 seconds. TOO
observations represents however a minor perturbation of the
schedule, with a 10\% probability of occurrence in the case of
BATSE \cite{Fish89}, and, as of 2003, SWIFT \cite{Gehrels00}, and
much less for HETE-2 \cite{Ricker01}.

The Telescope Control System is in charge of all the housekeeping
of the instrument, and actually sends the orders to the
telescope, auxiliary equipement, and triggers the Camera. In case
of bad weather conditions (high wind, rain, temperatures too
close to the dew point, etc.), the roof is closed, as it is the
case if the electric supply fails. In the other cases, the
operations are permitted, while eventually the processing
software may reject certain frames, and their corresponding
observation blocks (see below).

Because of the large pixel size (3.6 arcsec), the relatively high
sky background over the Calern, and the absence of guiding, the
length of a single exposure has to be limited to 5 minutes. In
practice, burst frames last 20 seconds to one minute, and most of
the other observations last less than two minutes. This
represents a major difference with the framework of large
telescopes, especially if we consider spectroscopic observations.
However, we consider that the generality of the algorithms
presented here allows the adaptation of the solution to longer
exposures. This is one of the goals of our present work on the
second version of the MAJORDOME software.

\section{The MAJORDOME paradigm}

\subsection{Types of request and constraints}

In the remaining of this paper we call a {\it request} a set of
observations requested by a given user. They may be not
contiguous. An {\it observation block} is a set of contiguous
images scheduled together.Though this is the general case, the
individual observations of an {\it observation block} may point to
different areas of the sky, as it is the case for mosaic pointings
generated from a BATSE GRB alert, or in the case of alternative
pointing between the target and calibration frames. The {\it
observation block} is the unit of programmation and of
validation. Each individual frame is defined by its duration, CCD
configuration, filter... Finally, as explained above, the
observation block is composed of individual {\it frames}.

In our approach any user can send a request at any time.
 These requests will be taken into account by the program at the next
  MAJORDOME program restart, or at a pre-established time,
  usually during daytime, or after any interruption.
 The lifetime of a request is one year from the submission date.
 This means that a request has no pre-allocated time
 (at least in the absence of any constraint),
 the limitation of 1 year being there only
 to avoid that a request not processed for
 any reason during this year remains in the database
 {\it ad vitam aeternam}. Some observational constraints
 may be affected to this request,
 like the Moon, etc.. There are 5 types of observations:

\begin{itemize}

\item The user can request a "constrained observation" (CO),
i.e.  it should
 be scheduled only during a given time interval.
 In the case of TAROT, it can be either minutes,
 days, or even months.
 The requested time, as
 well as the flexibility around this time should be given.

\item The request can call for "periodic not constrained"
(PNCO) observations, i.e. regularly spaced observations.
The period, as well as its uncertainty should be given
at request time. This period has no special limits,
i.e. it can be minutes (and even less), days or more.
The program has the freedom to schedule the
observations at any time, provided that all
observations can fit the night, for short periodic observations.

\item In the case of "periodic constrained"
observations (PCO), the user gives the
period and also the phase (the constraint
on the date of the first observation).

\item Finally an observation may not be "constrained"
nor "periodic" at all. In this case, we call it a
"non constrained observation" (NCO).

\item A somewhat specific type of "alert observations" (AO)
has been added. In the case of TAROT,
requests for observations of this type
are sent directly to the MAJORDOME by the
GCN, using a socket type INTERNET connection.
The transfer time between the NASA GSFC and the
 CESR as measured from several months of observations is about
250 ms. If the alert comes from a BATSE trigger, then the
uncertainty of the localization is larger than the field of view,
and a mosaic {\it observation block} is generated. Usually TAROT
spends the remaining of the night observing the burst location,
and eventually part of the following nights. This means that
alerts are unpredictable events with an impact on the overall
telescope schedule. Alert observations occupy about 10\% of the
telescope time. Though the alert processing is somewhat specific
to TAROT, it has a lot of analogies with less atypical Target
Opportunity Observations (TOOs) on classical telescopes.

\end{itemize}

The specific limits of TAROT are at present that an {\it
observation block} should have at most 6 individual observations
of duration less than 5 minutes.  As explained above, this last
limit is due to the high sky background of the Calern
observatory. Also, given that the MAJORDOME scheduling horizon is
at present one night, larger blocks may produce a loss of
efficiency. In the specific case of TAROT, this is not a problem
since this telescope is mainly devoted to the observation of
transient and variable objects, and that deep sky exposures are
beyond its scope. We note also that the unit of validation is the
{\it observation block}, i.e. if any of the individual observation
of a given observation block is not performed or considered as
valid (e.g. if they are clouds on one of the frames), then all the
{\it block} observations should be re-scheduled.

Finally there are four priority levels: periodic or
constrained observations are in the highest level, and we
defined three levels for NCO observations. The sum of the
three first levels (the periodic and/or constrained
observation levels and the two first level for NCOs)
should be close to 100\% of the foreseen observing time.
The third level of NCOs will be scheduled only as "filling"
observations if there is some remaining time, or if no
other higher priority observation can be scheduled at
that time. Of course, the amount of periodic or
constrained observations should be limited to a small
fraction (on the order of 20\%), in order to control the
distribution of the types of scheduled observations.

\subsection{Optimization quality}

The quality of a given timeline solution for any telescope
should be quantified. Here we describe the measures we
used:

\begin{itemize}

\item The efficiency of the system, i.e. the ratio
of the effective observing time (including "bad frames" due
to external events and the readout dead time) over the total
operational night time should be as high as possible.

\item The high priority levels should be served first.

\item The number of observations scheduled
close to their transit should be as high as possible.

\item The probability to schedule
short and long observations should be the same, e.g. the
MAJORDOME should not introduce a "duration preference".

\item The program should not give any preference
to any part of the sky, i.e. it should respect
the parity given in the request database
(i.e. no North or South preference).

\item The telescope moves should be minimized.

\end{itemize}

It has to be noted that several of these measures could be
used also in the decision process for a given timeline. However
the quality measure is done after the telescope has
performed the observations. Also, as described below,
there are other parameters which have an impact on the
computation of the optimized solution:

\begin{itemize}

\item {CO and PCO requests} : Since these requests have to be
scheduled at a precise date, their insertion in the
telescope timeline will occur as soon as possible and
don't have a specific criterion of quality.

\item {NCO requests} : These request should be
observed close to their transit.

\item{PNCO requests}: Same as above, excepted that as soon as one
of the observations is scheduled, the others become "constrained"

\end{itemize}
\section{The MAJORDOME software}

As explained in section 2 the MAJORDOME is the software that takes
control over the Telescope Control System (TCS) during night
operations, i.e. at night when no other event prevents the
telescope from observing (e.g. bad weather).


\subsection{connection with other modules}

The OBSERVATION REQUESTOR is a module that enables the users to
send a request to the telescope. The observation requests are
then put in the request database. Just before night time, or
after any interruption (weather, alert...) the MAJORDOME scans
this database and builds a timeline for the night, or the
remaining of it. At night time, in the absence of any other event
(e.g. rain) the MAJORDOME sends the observation orders to the
Telescope Control System (TCS). Whenever the frame has been
downloaded from the CAMERA, it is analyzed by our automatic
TAITAR data processing software \cite {Bringer99}. If the {\it
observation block} has been successfully observed, the
information is sent to the MAJORDOME, and the corresponding
request is removed from the request database as soon as the
observation block has been completely observed . All links
between the various software modules of TAROT are established
using a socket type connection.

\subsection{optimization}

The scheduling program is divided in 2 main parts :

\begin{itemize}[$\bullet$]
\item   the {\it selection} procedure.

\item   the {\it run} procedure.
\end{itemize}

The {\it selection procedure} selects from the request
database all the visible ones for the period of time where
the scheduling is needed. (the next night for a normal
procedure, or part of the night in case of rescheduling
during the night, the remaining of it if the operations
resume after an interruption). This selection is based on
the following tests:

\begin{itemize}
\item Minimal visibility of the object.
An observation can be made only if the object is at that
time above the walls of the TAROT building.  This means
that the declination should be above than
$-22^{\circ}$, and a constraint on the rising
and setting time of the object.

\item Night Observation: The object has to be visible
between astronomical twilight and dawn. However {\it alert}
observations may be scheduled during nautical dawn or
twilight. Other observation types may be scheduled also
during these intervals, like satellite or comet searches.

\item Minimal Moon distance: When processed,
the object has to be at least $10^{\circ}$ away from the Moon.

\end{itemize}

\begin{figure}[h!]
\centerline{\epsfig{file=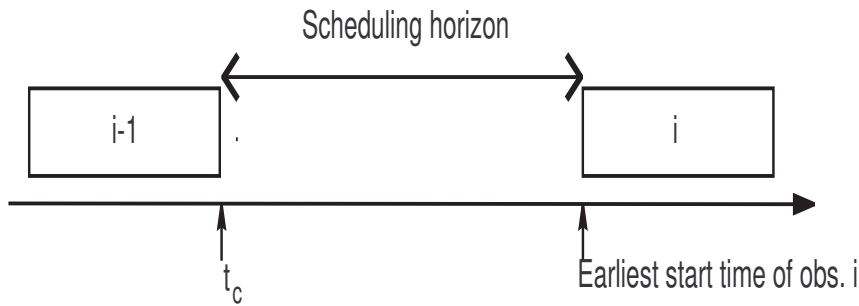,width=12.0cm,height=5.0cm}}
  \caption{Scheduling horizon. t$_{\rm c}$ is the "running time" (see text)}
  \end{figure}

We define the "night window" of a request as the time during which
it meets all the above criteria (visibility, Moon constraint,
etc.) for a given night. The "observation window" of a request is
the union of all the night window of a request during its
validity period (one year or less). The procedure produces a
subset of the database made of the requests which can be
scheduled, because the intersection of their "night windows" and
the actual night is non null.

The {\it run procedure} schedules a subset of requests selected by the
{\it selection procedure}. It does its job in two steps: first, CO,
and PCO requests are scheduled, and then it places the PNCO and
NCO requests. We call {\it scheduling horizon} the non null time interval
between the end and the beginning of two observations (figure 2).

\subsubsection{ CO and PCO scheduling }

Observations are ordered by their requested date. The first one is
placed at its earliest starting time. The running date,
$\rm{t}_{\rm{c}}$ becomes the date of the first observation plus
its duration. Then, for every observation i in the list ($\rm{i}
= 1 , 2 \ldots n$):

\begin{description}
    \item{Case 1:} No overlapping (figure 2): observation i is scheduled
    at its earliest starting time, and the running date $\rm{t}_{\rm{c}}$
    becomes the current starting time of i plus its duration.

    \item{case 2:} Overlapping (figure 3): If it is possible to delay observation
    i, or to invert the occurrence of i and i-1,
    taking into account their tolerance on the starting date and the end of their night window,
    then i is scheduled at the running date $\rm{t}_{\rm{c}}$ (figure 3a), and
    the running date becomes $\rm{t}_{\rm{c}}$ plus the duration of observation
    i (or i-1). Else, observation i is rejected (figure 3b), even if it is one of the scheduled
    occurrence of a  PCO observation.

\end{description}

  \begin{figure}[h!]
\centerline{\psfig{file=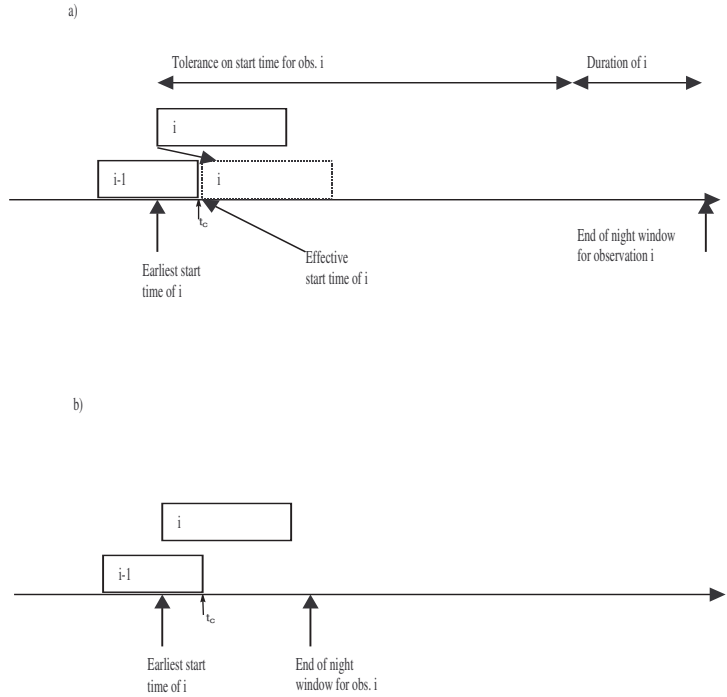,width=10.0cm,height=10.0cm}}
  \caption{Two cases of overlapping observations: a) there is
  a possibility to delay i, b) i cannot be delayed, hence this
  observation will not be scheduled}
  \end{figure}

\subsubsection{PNCO scheduling}

At present for PNCO observations, which encompass usually the
whole night, we sequence the first observation as soon as
possible, after PCO and constrained observation have been placed.
To do that we follow the above defined rules. As soon as the
first observation of the periodic sequence has been scheduled,
the others become {\it de facto} constrained, and they are
sequenced according to the rules defined for PCO observations.

Though this placement is not satisfactory from the principles, in
fact, for periods shorter than 3 hours, there is always an
observation sequenced near its transit time. However, in order to
get a better optimized version of our method, in the next version
we will implement a global solution which will enable a better
placement of PNCO observations, especially when they run on the
long time, or, on the contrary, when the total duration
encompassed by the observations is short compared to the night
length.

\subsubsection{NCO scheduling}

During this step we will try to schedule the NCO observations
within the scheduling horizons (see above).We use the pairwise
interchange algorithm\cite{Baker79} between observation tasks, in
order to minimize an evolution function taking into account the
airmass. By scheduling CO and PCO, we have defined several {\it
scheduling horizons}. NCO and PNCO have to be scheduled within
these intervals. Let us coinsider a given observation I
associated with a NCO request (figure 4):

\begin{itemize}

\item $tr_{\rm I}$ is the transit time of I.
\item $t_{\rm dI}$ is the instant of the beginning of observation I.
\item $d_{\rm I}$ is the duration of observation I.
\item $d_{\rm tI}$ is a measure of the distance between
the transit and scheduling times.

\end{itemize}

 \begin{figure}[h!]
\centerline{\epsfig{file=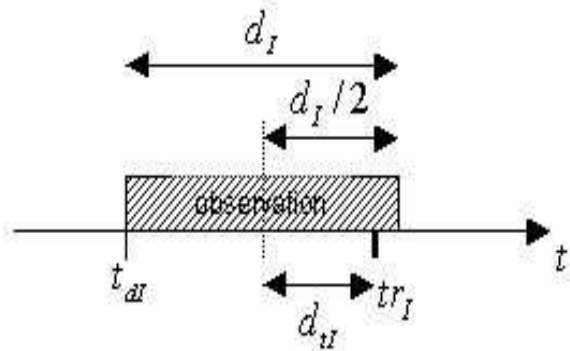,width=9.0cm,height=5.0cm}}
  \caption{Parameters associated with a non constrained observation}
  \end{figure}

We use as the measure for $d_{\rm tI}$:

\begin{equation}
 d_{\rm tI} = tr_{\rm I} -(t_{\rm dI} + d_{\rm I}/2)
\end{equation}

We will adopt $\bar{d}$, the mean distance to transit as a
quality criterion of a solution where :

\begin{equation}
\bar{d} = 1/N[\sum_{I=0}^{N}d_{\rm tI}]   
\end{equation}

Here, N is the number of observations.

For every scheduling horizon, the corresponding timeline is built
progressively. We take the best scheduling of two successive
observations, taking into account their characteristics (here the
distance to the transit).

a) Best scheduling of two consecutive observations

Let us now consider two NCOs A and B and a current date $t_c$. We
have for $t_c$, the choice to schedule A-B or B-A (figure 5).

  \begin{figure}[h!]
\centerline{\epsfig{file=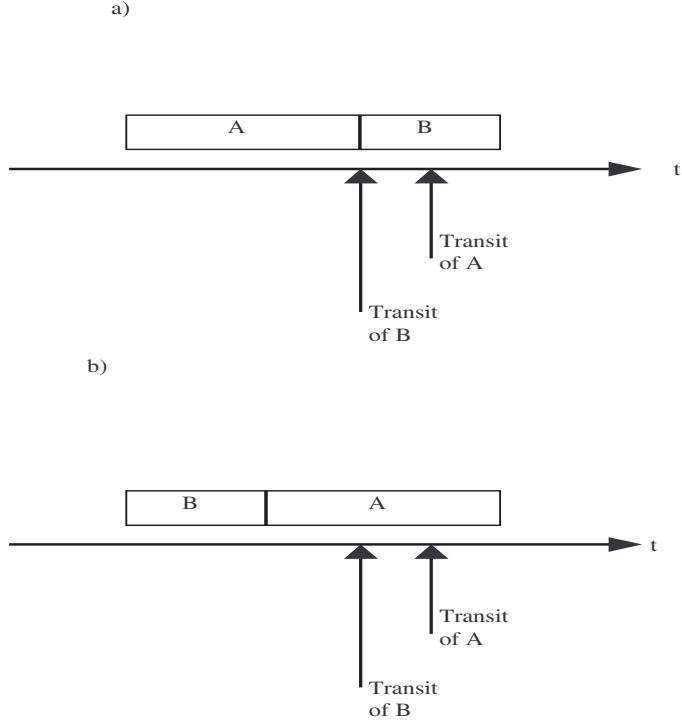,width=10.0cm,height=10.0cm}}
  \caption{A has greater priority than B, but the better sequence
  let include B before A in the timeline because of the
  respective transit times}
  \end{figure}

\begin{itemize}[$\bullet$]

\item Case 1 : A and B have different priorities:
when two NCO have different priorities, the higher priority
holder imposes the sequence that minimizes its distance to
transit.

\item Case 2 : A and B have the same priorities,
but the number of possible remaining transits (up to the end of
the observing window)  for A is greater than for B. In that case,
B has higher priority over A.

\item Case 3 : Same priorities and same number of transits:
When two observations are equivalent, we choose the sequence that
minimizes the contribution to $\bar{d}$. We do not compute
$\bar{d}$, but we compare the following numbers $C_{\rm AB}$ and
$C_{\rm BA}$, where $C_{\rm AB}$ is the sum of distances to
transit of the observations A and B when A is programmed in first
place ($ t_{\rm dA}=t_{\rm c}$ and B is scheduled in second
position ($t_{\rm dB} = t_{\rm c}+ d_{\rm A}$) :

\end{itemize}

\begin{equation}
[C_{\rm AB} = |tr_{\rm A} - (t_{\rm c}+ d_{\rm A}/2)| + |tr_{\rm
B} - (t_{\rm c} +d_{\rm A} + d_{\rm B}/2)|]
\end{equation}

\begin{equation}
[C_{\rm BA} = |tr_{\rm B} - (t_{\rm c}+d_{\rm B}/2)| + |tr_{\rm
A} - (t_{\rm c}+d_{\rm B} + d_{\rm A}/2)|]
\end{equation}

if $C_{\rm AB} < C_{\rm BA}$, then we choose the A-B sequence,
otherwise the B-A one.

b) Global scheduling algorithm within the scheduling horizons

Observations are ordered by growing distance to the transit. We
consider the scheduling horizons in the chronological order. For
a given horizon H:

\begin{itemize}
\item We place the first observation of the list whose duration is
consistent with H (if this is not possible we consider the next
horizon).

\item We fill H adding progressively the observation compatible with
the remaining duration. Any new observation is scheduled against
the preceding one according to the rule developed in \S a). When
no new observation can be introduced in the remaining duration
interval, then we consider the next horizon.

\end{itemize}

\subsection {Alert (TOO) observations}

Should an alert occurs, e.g. from the HETE-2 satellite, a new
timeline is immediately built for the remaining of the night with
only the alert observations. The MAJORDOME sends to the TCS the
first alert observation. This procedure takes less than one
second. A new PCO request is built and inserted in the request
database, for follow-up observations beginning the following
night.

\subsection {Computing requirements}

Since TAROT has a somewhat high rate of interruption (30\% of
them due to bad weather and 10\% because of a GCN alert), our
algorithm should not require too much CPU. We have tested the
MAJORDOME on a 233 MHz PC Pentium II with 128 MB of RAM. With the
present TAROT camera it takes less than 600 seconds to establish a
timetable for the night, with a database of 1500 requests of
different types, and 300 to 400 images schedules (about 100
observation blocks). The main parameters which have an impact on
the computing time are the size of the request database, and the
number of observation blocks it is possible to schedule. This
last parameters depends on the length of the night, and more
generally of the horizon (i.e. the larger the horizon, the larger
the number of possible combinations), and, at least for TAROT,
the total time taken for each exposure: for TAROT, this last
parameter was dramatically reduced when we replaced the camera,
since the readout time was reduced from 30 seconds to 2 seconds.

\section{results}
We have tested our method on both real and simulated request
databases. Table 1 summarizes the characteristics and the results
for the scheduling with the simulated request database.:

\begin{table}
\caption{Characteristics and results obtained with the
simulated request database}
\begin{tabular}{lllll}\hline
Request type&Number of requests&Total time &Scheduled
requests&Total
time \\
  & & (min) & & (scheduled requests)\\
   \hline
Total&500&1476.38& & \\
Nigh duration& &379& &379\\
Observable requests&435&1301.38&121&338.85\\
PNCO&41&233.20&12&46.01\\
NCO&359&974.01&74&190.67\\
CO + PCO&35&94.17&35&94.17\\
\hline
\end{tabular}
\end{table}

\subsection{efficiency of the system}
The efficiency $\epsilon$ of the system is simply defined
as the ratio of the effective observing time by the total
operational night time.

$\epsilon=$cumulated observing time / total operational
night time

The above database leads to  $\epsilon = 88.59\%$. As it
can be naturally inferred, the efficiency of the schedule
is closely related to the number and distribution of CO,
PCO and even  PNCO requests. In the above example, the PCO
+ CO requests represent about 28\% of the total night
duration. If we add the PNCO, more than 40\% percent of
the observations have a constrain of any type. If this
number is lowered to about 10 - 15\%, then the efficiency
grows up to 90 or 95\%.  Since the algorithm gives a local
solution and tries first to schedule requests at transit
time, the consequence is a loss of efficiency when too many
requests are near the same right ascension. We did not
observe however any North-South preference, or even
duration preference, since this last criterion is only
marginally taken into account.

\subsection{Scheduling  near transit}

The NCO are designed to be scheduled near their transit time with
a certain tolerance (60 minutes). On figure 6, we can clearly see
that this is indeed the case.

 Some NCO have to use their tolerance,
 depending on the number of requests already
 scheduled and on the distribution of COs and PCOs previously scheduled.

\section{Discussion and conclusions}

The MAJORDOME is now routinely operating  the TAROT instrument.
The advantage of this software is its ability to cope with various
situations, while ensuring an efficient scheduling and optimal
observations.

We have identified two main ways to improve the MAJORDOME: First
the present MAJORDOME works on a local solution. A more global
approach would be desirable both for the efficiency of the
scheduling, to allow observations having the same transit time to
be adequately scheduled over a longer period, to better manage
priorities based on the length of the celestial window. Another
advantage of long-term scheduling is that the user may know in
advance the approximate date of her/his observation, with a
certain level of confidence. In order to overcome this difficulty
we are working on long-term request pre-allocation (giving an
approximate date of consideration for the scheduler), and
mid-term scheduling. Request pre-allocation is also a mean to
give a better balance between short and long duration
observations. Of course, these efforts should not be made at the
expanse of the reactivity in case of alerts or bad weather, nor
prevent the insertion of new requests at any time, particularly
when a TOO occurs.

On the other hand one may be tempted to use the same system to
compute the timeline for a longer period, i.e. to use the current
algorithm with a larger horizon. As it has been described above,
this solution will produce unacceptable computing times, and may
also lead to inefficiencies, since in the version described here
we scan the whole database each time the MAJORDOME starts the
computation of a new timeline. The solution we foresee would
benefit of the advantages of the current MAJORDOME, used with a
mid-term horizon (e.g. a week), and a pre-placement of the {\it
observation blocks} on the long term. This may also an advantage
for the scheduling of periodic observations when the period is
large.

 \begin{figure}[h!]
  \caption{Relative NCO placement with respect to their transit
  time}
  \end{figure}

Secondly, we are working on a more flexible MAJORDOME,
adding several criteria to the requests, like observations
at dark/grey/Moon time, between clouds (in case of partial
coverage), during photometric nights only... We are also
introducing other observation types, like repeated
observations, when the user needs only to compare regions
of the sky at different times, but with a soft constraint
on the observation dates.

We designed the MAJORDOME as a software whose goal was to
schedule requests not on the basis of a pre-allocated time, but
in an optimal way. In other words the user is only guaranteed
that if her/his observation has enough priority, it will be
observed in the best possible conditions, but he does not know
exactly when it will be observed, though we are trying now to
give a good guess of this date.

From the point of view of the ability to cope with
unexpected events, alerts or bad weather, the MAJORDOME
proved already its usefulness, since a new timeline is
computed in few seconds as a response of any interruption,
eventually inserting new, urgent observations.

\acknowledgements The T\'elescope \`a Action Rapide pour les
Objets Transitoires (TAROT-1) has been funded by the Institut
National des Sciences de l'Univers (INSU) of the Centre National
de la Recherche Scientifique (CNRS), and has been built by the
Centre d'Etude Spatiale des Rayonnements (CESR/CNRS), the
Laboratoire d'Astronomie Spatiale (LAS/CNRS), the Technical
Division of the Institut National des Sciences de l'Univers
(INSU-DT), and is hosted by the Observatoire de la C\^ote d'Azur
(OCA). We thank the anonymous referee for her/his helpful comments

\end{article}

\end{document}